\documentclass[floatfix,aps,amsmath,nofootinbib,onecolumn,10pt]{revtex4}

\usepackage{listings}
\usepackage{graphicx}
\usepackage{bm}
\usepackage{rotating}
\usepackage{array}
\usepackage{amsmath}
\usepackage{amssymb} 
\usepackage{mathrsfs} 
\usepackage{cancel}

\def\half{{1\over 2}}

\def\half{{1\over 2}}
\def\({\left(}
\def\){\right)}
\def\[{\left[}
\def\]{\right]}

\def\e{\begin{equation}}
\def\q{\end{equation}}
\def\m{\begin{eqnarray}}
\def\n{\end{eqnarray}}

\begin{document}

\title{Distance Priors from Planck Final Release}

\author{ Lu Chen$^{1,2}$\footnote{chenlu@itp.ac.cn}, Qing-Guo Huang$^{1,2,3,4}$ \footnote{huangqg@itp.ac.cn} and Ke Wang$^{5}$ \footnote{wangke@itp.ac.cn}}
\affiliation{$^1$ CAS Key Laboratory of Theoretical Physics,\\ Institute of Theoretical Physics, \\Chinese Academy of Sciences, Beijing 100190, China\\
$^2$ School of Physical Sciences, \\University of Chinese Academy of Sciences,\\ No. 19A Yuquan Road, Beijing 100049, China\\
$^3$ Synergetic Innovation Center for Quantum Effects and Applications, Hunan Normal University, 36 Lushan Lu, 410081, Changsha, China\\
$^4$ Center for Gravitation and cosmology, College of Physical Science and Technology, Yangzhou University, 88 South University Ave., 225009, Yangzhou, China\\
$^5$ National Astronomical Observatories, \\Chinese Academy of Sciences,  20A Datun Road, Beijing 100012, China\\}

\date{\today}

\begin{abstract}

We present the distance priors from the finally released $Planck~ \textrm{TT,TE,EE}+\textrm{lowE}$ data in 2018. The uncertainties are around $40\%$ smaller than those from $Planck$ 2015 TT$+$lowP.  In order to check the validity of these new distance priors, we adopt the distance priors to constrain the cosmological parameters in different dark energy models, including the $\Lambda$CDM model, the $w$CDM model and the CPL model, and conclude that the distance priors provide consistent constraints on the relevant cosmological parameters compared to those from the full $Planck$ 2018 data release.
 
\end{abstract}

\pacs{???}

\maketitle


\section{Introduction}
\label{introduction}

Since two supernova surveys reported the discovery of cosmic acceleration independently in 1998 \cite{Riess:1998cb,Perlmutter:1998np}, a new component except from matter and radiation, named dark enenrgy (DE) \cite{Peebles:2002gy}, is required assuming the general relativity (GR) remains correct for our universe. 
DE is the mathematically simplest explanation to the accelerating expansion of the universe, but its nature is still a puzzle. 
Several methods can be used to give constraints on the properties of DE. 
A straight-forward method is the distance measurement, such as the direct determination of $H_0$ \cite{Riess:2016jrr}, surveys on Type Ia Supernovae (SNe) \cite{Conley:2011ku,Suzuki:2011hu} and the baryon acoustic oscillation (BAO) measurements \cite{Cole:2005sx}. They provide absolute or relative distance measurements in a narrow range of redshift with percent level uncertainties. Obviously, the narrow detecting range of redshift restricts the validity of exploring the evolution of DE in full range of redshift. Moreover, the uncertainties increase as the redshift gets higher. 
DE is a component with negative pressure, which produces a force of repulsion, and affects the galaxy clustering. Then we can explore it using gravitational lensing \cite{Sanchez:2006nj}, clusters of galaxies \cite{Neveu:2017jkg}, redshift-space distortions (RSD) \cite{Gil-Marin:2015sqa} and the Alcock-Paczynksi (AP) effect \cite{Alcock:1979mp}. However, our limited knowledge about the structure formation bring a big challenge in the specific processing. 
We can also use the cosmic microwave background (CMB) \cite{Komatsu:2008hk,Aghanim:2018eyx} to constrain DE properties because DE plays an important role in the matter constitution and leaves footprints on the late-time power spectra of CMB. This method breaks the limitation of redshifts and possesses high-accuracy. Combining other measurements, a spatially flat $\Lambda$CDM model remains a convincing model. But it also has limitations that it requires the full Boltzmann analysis \cite{Bean:2003fb,Weller:2003hw,Li:2008cj}, which is really a very time-consuming process. More importantly, equations for linear density perturbations in some DE models are difficult to build \cite{Dvali:2000xg,Koyama:2006ef}. As a result, the method of distance priors \cite{Bond:1997wr,Efstathiou:1998xx,Wang:2007mza} are proposed to be a compressed likelihood to substitute the full Boltzmann analysis of CMB. 
 
Since the first data release in 2013 \cite{Ade:2013zuv}, Planck satellite provides CMB data with high accuracy. Although the preliminary observations of $TE, EE$ power spectrum at high multipoles were released in $Planck$ 2015 \cite{Ade:2015xua}, this data release laid emphasis on the temperature power spectrum.
Recently, Planck Collaboration release the final data of the CMB anisotropies (hereafter $Planck$ 2018) \cite{Aghanim:2018eyx}. 
Since improved measurements of low-$l$ polarization allow the reionization optical depth to be determined with higher precision compared to $Planck$ 2015, there are significant gains in the precision of some parameters which are correlated with the reionization optical depth. Due to improved modelling of the high-$l$ polarization, moreover, there are more robust constraints on many parameters which will be affacted by residual modelling uncertainties only at the $0.5\sigma$ level.
The constraints on the distance priors given in ``Planck Blue Book" \cite{Planck:2006aa} are  about $50\%$ smaller than those given by $Planck$ 2015 TT$+$lowP \cite{Ade:2015rim}.
All in all, it is meaningful to update the distance priors with the full-mission $Planck$ measurement of CMB.

Following the previous work in \cite{Huang:2015vpa}, we update the distance priors with $Planck$ 2018 and present the constraints on several DE models with these new distance priors. This paper is organized as follows. In Sec.\ref{method},  we show our methodology to reconstruct the distance priors from $Planck$ 2018 chains. Then the new distance priors are  presented in Sec.\ref{results}. In Sec.\ref{de}, we check our results in several different DE models. Concretely, we constrain the equation of state of DE from distance priors and compare our results with those by fitting the full data of $Planck$ 2018 release. A brief summary is given in Sec.\ref{sum}. In addition, we provide a note on how to use the distance priors in the CosmoMC package in the appendix which should be quite useful for the readers. 

\section{Distance priors from Planck 2018 data and constraints on DE models}
\label{body}

\subsection{Methodology}
\label{method}
The distance priors provide effective information of CMB power spectrum in two aspects: the acoustic scale $l_\textrm{A}$ characterizes the CMB temperature power spectrum in the transverse direction, leading to the variation of the peak spacing, and the ``shift parameter" $R$ influences the CMB temperature spectrum along the line-of-sight direction, affecting the heights of the peaks.
 
We adopt the popular definitions of the distance priors as follows \cite{Komatsu:2008hk}:
\begin{align}\label{la}
l_\textrm{A} =(1+z_*)\frac{\pi D_\textrm{A}(z_*)}{r_s(z_*)}\ ,
\end{align}
\begin{align}\label{Rz}
R(z_*)\equiv\frac{(1+z_*)D_\textrm{A}(z_*) \sqrt{\Omega_m H_0^2}}{c}, 
\end{align}
where $z_*$ is the redshift at the photon decoupling epoch. Here we use the values of $z_*$ given by the $Planck$ 2018 chains. $r_s$ is the comoving sound horizon, defined by
\begin{align}\label{}
r_s(z)  &= \frac{c}{H_0} \int_0^{1/(1+z)} \frac {da} {a^2E(a)\sqrt{3(1+\frac{3\Omega_bh^2}{4\Omega_\gamma h^2}a)}}\ ,\nonumber\\
\frac{3}{4\Omega_\gamma h^2}  &= 31500(T_{\textrm{CMB}}/2.7K)^{-4},~~T_{\textrm{CMB}}= 2.7255K\ .
\end{align}
And the angular diameter distance $D_\textrm{A}$ is given by
\begin{align}\label{}
D_\textrm{A}
=\frac{c}{(1+z) H_0 \sqrt{|\Omega_{k}|}  } \textrm{sinn}\left[|\Omega_{k}|^{1/2}\int_0^z \frac {dz'} {E(z')}\right]\ ,
\end{align}
where $\textrm{sinn}(x) \equiv \textrm{sin}(x)$, $x$, $\textrm{sinh}(x)$ for $\Omega_{k}<0$, $\Omega_{k}=0$, $\Omega_{k}>0$ respectively. Here $E(z)$ is $E(z) \equiv{ H(z)}/{H_0}$, i.e.
\begin{align}\label{}
E(z) &= \[ \Omega_{r}(1+z)^4+\Omega_m(1+z)^3+\Omega_{k}(1+z)^2+\Omega_{de} \frac {\rho_{de}(z)} {\rho_{de}(0)} \]^\half\ ,
\end{align}
where $\Omega_{r}$ is the present fractional radiation density
\begin{align}\label{}
\Omega_{r}=\frac{\Omega_{m}}{1+z_{\textrm{eq}}}, ~~z_{\textrm{eq}}=2.5\times10^{4}\Omega_{m}h^2\left(T_{CMB}/2.7\textrm{K}\right)^{-4}\ .
\end{align}
For $\Lambda$CDM and $w$CDM models where $w$ is a constant, $ {\rho_{de}(z)} /{\rho_{de}(0)}$ equals $1$ and $(1+z)^{3(1+w)}$, respectively.
 
\subsection{Results}
\label{results}

In this section, we derive the distance priors in several different models using $Planck$ 2018 TT,TE,EE $+$ lowE which is the latest CMB data from the final full-mission Planck  measurement \cite{Aghanim:2018eyx}. We get the chains of $ l_\textrm{A}$ and $R$ from the public Planck chains\footnote{From the $Planck$ 2018 release, four sets of chains named as 
base\_plikHM\_TTTEEE\_low$l$\_lowE, base\_w\_plikHM\_TTTEEE\_low$l$\_lowE, \\
base\_omegak\_plikHM\_TTTEEE\_low$l$\_lowE and base\_Alens\_plikHM\_TTTEEE\_low$l$\_lowE are used in our paper to generate Table.\ref{distance priors}. \\
 And chains named  
base\_plikHM\_TTTEEE\_low$l$\_lowE, base\_w\_plikHM\_TTTEEE\_low$l$\_lowE\_BAO and \\
base\_w\_wa\_plikHM\_TTTEEE\_low$l$\_lowE\_BAO are used to generate the contours in Sec.\ref{de}.} under the corresponding models released on 17 July 2018 with Eqs.~(\ref{la}) and (\ref{Rz}), then marginalize over the parameters except for $\{R, l_\textrm{A}, \Omega_b h^2, n_s\}$. Distance priors are usually used to research the late-time univese expansion, so we present $\Omega_b h^2$ too. The scalar spectral index $n_s$ is shown for the convenience of studying the matter power spectrum.

\begin{table*}[!htp]
\centering
\renewcommand{\arraystretch}{1.5}
\begin{tabular}{cccccc}
\hline\hline
  $\Lambda$CDM & $Planck~ \textrm{TT,TE,EE}+\textrm{lowE} $ & $R$ & $l_\textrm{A}$ & $\Omega_b h^2$ &$n_s$ \\
\hline 
$R$                                           & $1.7502\pm0.0046$       & $1.0$&    $0.46$&   $-0.66$&   $-0.74$ \\
$l_\textrm{A}$                                & $301.471^{+0.089}_{-0.090}$        & $0.46$&    $1.0$&   $-0.33$&   $-0.35$ \\
$\Omega_b h^2$                              & $0.02236\pm0.00015$      & $-0.66$&   $-0.33$&  $1.0$&    $0.46$\\
$n_s$                                        & $0.9649\pm0.0043$       & $-0.74$&   $-0.35$&  $0.46$&    $1.0$\\
\hline\hline
  $w$CDM & $Planck~ \textrm{TT,TE,EE}+\textrm{lowE} $ & $R$ & $l_\textrm{A}$ & $\Omega_b h^2$ &$n_s$ \\
\hline 
$R$                                           & $1.7493^{+0.0046}_{-0.0047}$       & $1.0$&    $0.47$&   $-0.66$&   $-0.71$ \\
$l_\textrm{A}$                                & $301.462^{+0.089}_{-0.090}$         & $0.47$&    $1.0$&   $-0.34$&   $-0.36$ \\
$\Omega_b h^2$                               & $0.02239\pm0.00015$      & $-0.66$&   $-0.34$&  $1.0$&    $0.44$ \\
$n_s$                                        & $0.9653^{+0.0043}_{-0.0044}$       & $-0.72$&   $-0.36$&  $0.44$&    $1.0$\\
\hline\hline
   $\Lambda$CDM$+\Omega_k$ & $Planck~ \textrm{TT,TE,EE}+\textrm{lowE} $ & $R$ & $l_\textrm{A}$ & $\Omega_b h^2$ &$n_s$\\
\hline 
$R$                                           & $1.7429\pm0.0051$       & $1.0$&    $0.54$&   $-0.75$&   $-0.79$ \\
$l_\textrm{A}$                                & $301.409\pm0.091$        & $0.54$&    $1.0$&   $-0.42$&   $-0.43$\\
$\Omega_b h^2$                              & $0.02260\pm0.00017$      & $-0.75$&   $-0.42$&  $1.0$&    $0.59$ \\
$n_s$                                        & $0.9706^{+0.0047}_{-0.0050}$       & $-0.79$&   $-0.43$&  $0.59$&    $1.0$\\
\hline\hline
   $\Lambda$CDM$+A_\textrm{L}$ & $Planck~ \textrm{TT,TE,EE}+\textrm{lowE} $ & $R$ & $l_\textrm{A}$ & $\Omega_b h^2$ &$n_s$\\
\hline 
$R$                                           & $1.7428\pm0.0053$       & $1.0$&    $0.52$&   $-0.72$&   $-0.80$\\
$l_\textrm{A}$                                & $301.406^{+0.090}_{-0.089}$        & $0.52$&    $1.0$&   $-0.41$&   $-0.43$ \\
$\Omega_b h^2$                              & $0.02259\pm0.00017$      & $-0.72$&   $-0.41$&  $1.0$&    $0.58$ \\
$n_s$                                        & $0.9707\pm0.0048$       & $-0.80$&   $-0.43$&  $0.58$&    $1.0$ \\ 
\hline
\end{tabular}
\caption{The 68$\%$ C.L. limits for $R$, $l_\textrm{A}$, $\Omega_b h^2$ and $n_s$ in different cosmological models and their correlation matrix for  from $Planck\  2018~ \textrm{TT,TE,EE}+\textrm{lowE} $. Notice that the Planck Collaboration use $Planck$ TT,TE,EE$+$lowE to represent the combination of the combined likelihood of TT,TE,EE spectra at $l\geq$ 30, the low-$l$ temperature Commander likelihood and the low-$l$ SimAll EE likelihood \cite{Aghanim:2018eyx}.}
\label{distance priors}
\end{table*}

\begin{figure}[]
\begin{center}
\includegraphics[scale=0.35]{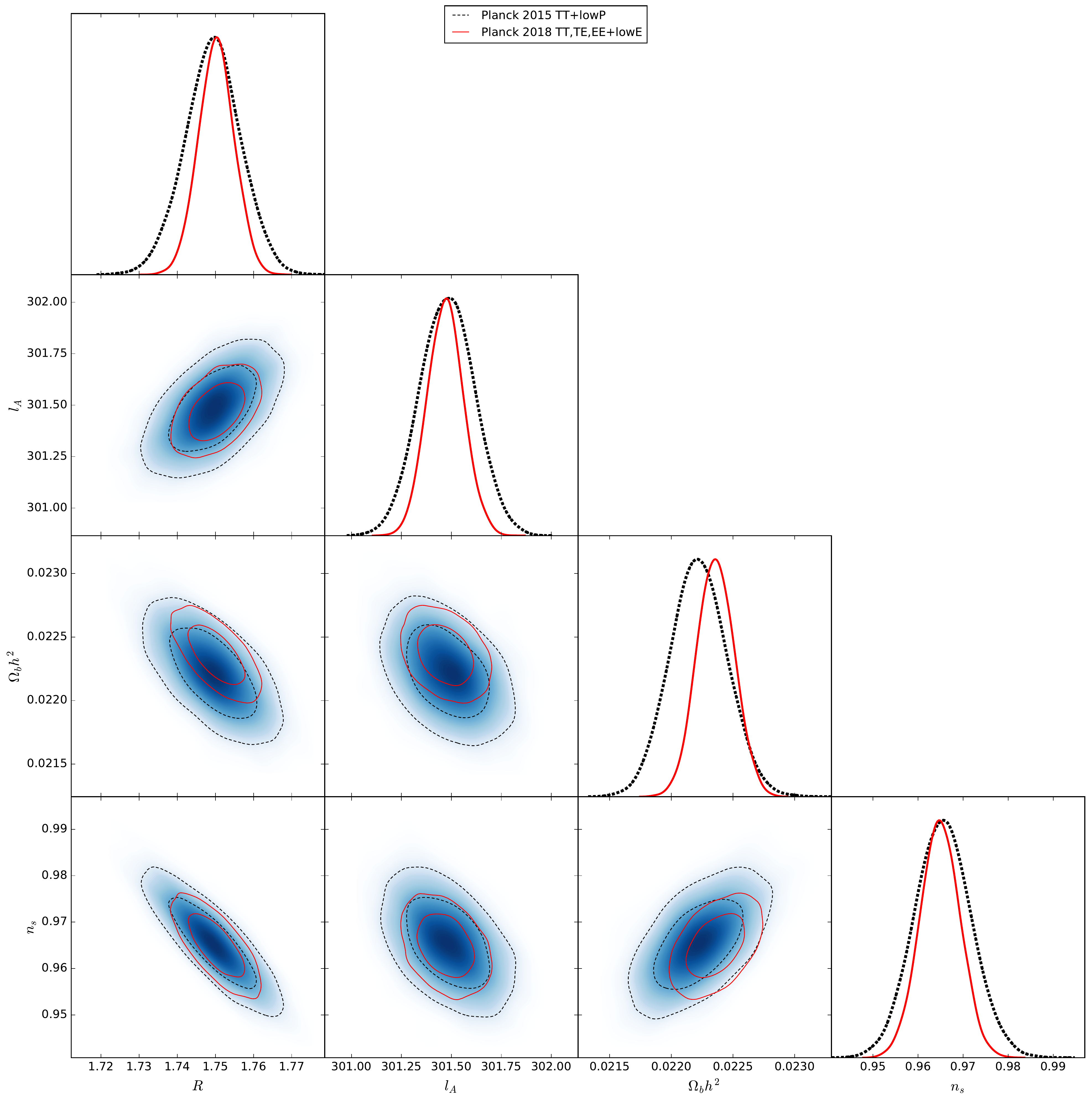}
\end{center}
\caption{Constraints on the distance priors in the base $\Lambda$CDM model from $Planck$ 2015 TT$+$lowP (the black, dashed contours) and $Planck$ 2018 TT,TE,EE$+$lowE (the red, solid ones).}
\label{fig:comparedistance}
\end{figure}

 The main results are shown in Table.~\ref{distance priors} where we list their 68$\%$ C.L. constraints and correlation matrices on the base $\Lambda$CDM model and its three 1-parameter extended models. From the results of the base $\Lambda$CDM model and the $w$CDM model, we can see that the distance priors are stable effective observables.  Considering that the geometric degeneracy can be broken up significantly by the smoothing effect of CMB lensing on the power spectrum, which is scaled by the weak lensing parameter $A_L$ \cite{Calabrese:2008rt}, we also provide the constraints on the $\Lambda$CDM$+\Omega_k$ model and the $\Lambda$CDM$+A_\textrm{L}$ model. Comparing to the previous two models, the restrictions on $R$ are weaken over $10 \%$ and those on $A_L$ are slightly weaken too, which is consistent with the theoretical prediction.

In Fig.~\ref{fig:comparedistance}, we compare the constraints on the distance priors $R$, $l_\textrm{A}$, as well as $\Omega_b h^2$ and $n_s$ in the base $\Lambda$CDM model derived from $Planck$ 2015 TT$+$lowP and $Planck$ 2018 TT,TE,EE$+$lowE. Obviously, the constraints from the $Planck$ final data release in 2018 are significantly improved. Actually, the errors from $Planck$ 2018 TT,TE,EE$+$lowE are around $40\%$ smaller than those from $Planck$ 2015 TT$+$lowP.

\subsection{Constraints on Dark Energy Models Using Distance Priors}
\label{de}
   
In this section, we use the distance priors derived for the base $\Lambda$CDM model in the former subsection to constrain the cosmological parameters in the base $\Lambda$CDM model, the $w$CDM model and the CPL model, and compare our results with the $Planck$ 2018 results to test the validity of the distance priors given in this paper. 

We modify the MCMC chains package CosmoMC \cite{Lewis:2002ah} by adding $\chi^2_{\rm distance\ priors}$, which is given by
 \begin{align}\label{}
\chi^2_{\rm distance\ priors}=\sum(x_i-d_i)(C^{-1})_{ij}(x_j-d_j),
\end{align}
where $x_i=\{R(z_*), l_\textrm{A}(z_*), \Omega_b h^2\}$ are values predicted in different DE models, $d_i=\{R^{Planck}, l_\textrm{A}^{Planck}, {\Omega_b h^2} ^{Planck}\}$ are set to their mean values and $C_{ij}$ is their correlation matrix in the $\Lambda$CDM model. $(C^{-1})_{ij}$ means its inverse matrix.
Here we use the approximate formula of $z_*$ to calculate $x_i$ \cite{Hu:1995en}
\e
z_* = 1048[1+0.00124(\Omega_b h^2)^{-0.738}][1+g_1(\Omega_m h^2)^{g_2}]\ ,
\q
where
\m
g_1 &=& \frac{0.0738(\Omega_b h^2)^{-0.238}}{1+39.5(\Omega_b h^2)^{0.763}}\ ,\\
g_2 &=& \frac{0.560}{1+21.1(\Omega_b h^2)^{1.81}}\ .
\n

\begin{figure}[]
\begin{center}
\includegraphics[scale=0.4]{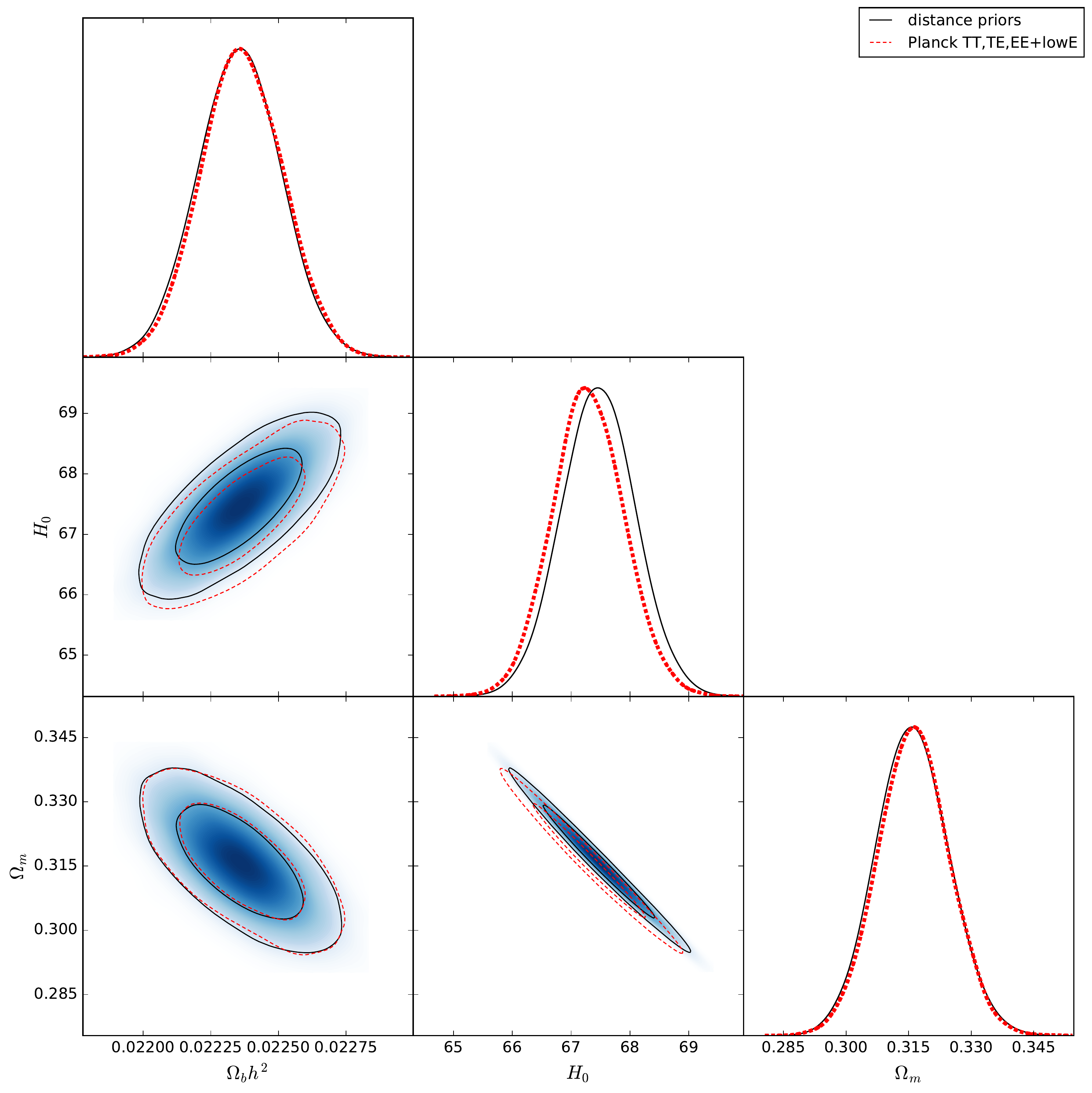}
\end{center}
\caption{Constraints on parameters in the base $\Lambda$CDM model. The $Planck$ 2018 chains gives the red dashed contours and the black solid ones are our results from distance priors for the base $\Lambda$CDM model.}
\label{fig:lcdm}
\end{figure}

In the base $\Lambda$CDM model, we constrain the set of parameters $\{\Omega_m, H_0, \Omega_b h^2\}$. In Fig.~\ref{fig:lcdm}, we show the comparison of our results from distance priors and the global fitting results from $Planck$ 2018. Clearly, the contours are almost overlaps, which indicates that the distance priors can take place of full $Planck$ released data effectively.

\begin{figure}[]
\begin{center}
\includegraphics[scale=0.35]{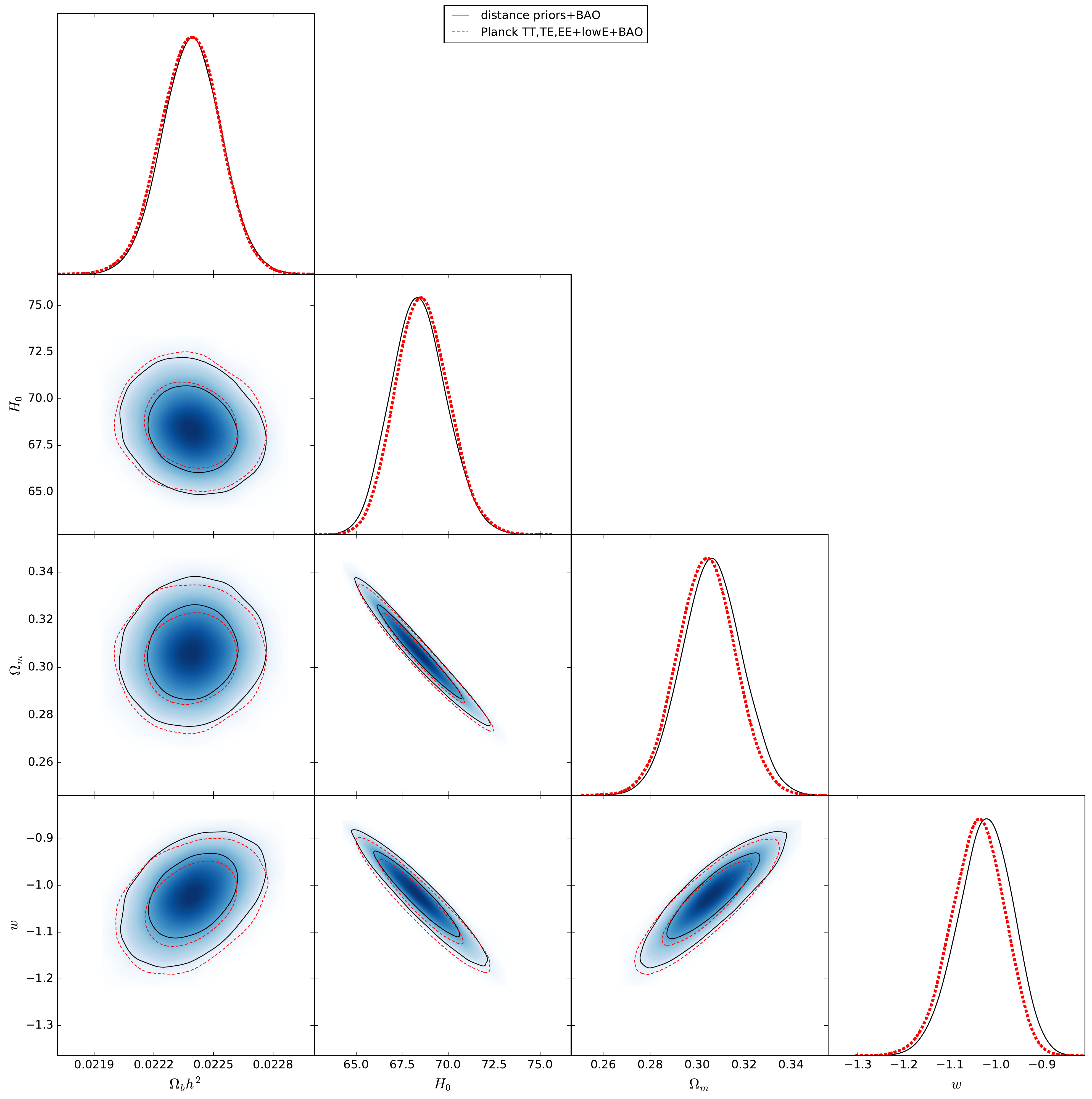}
\end{center}
\caption{Constraints on parameters in the $w$CDM model. The $Planck$ 2018 chains gives the red dashed contours and the black solid ones are our results from distance priors for the base $\Lambda$CDM model.}
\label{fig:w}
\end{figure}

\begin{figure}[]
\begin{center}
\includegraphics[scale=0.3]{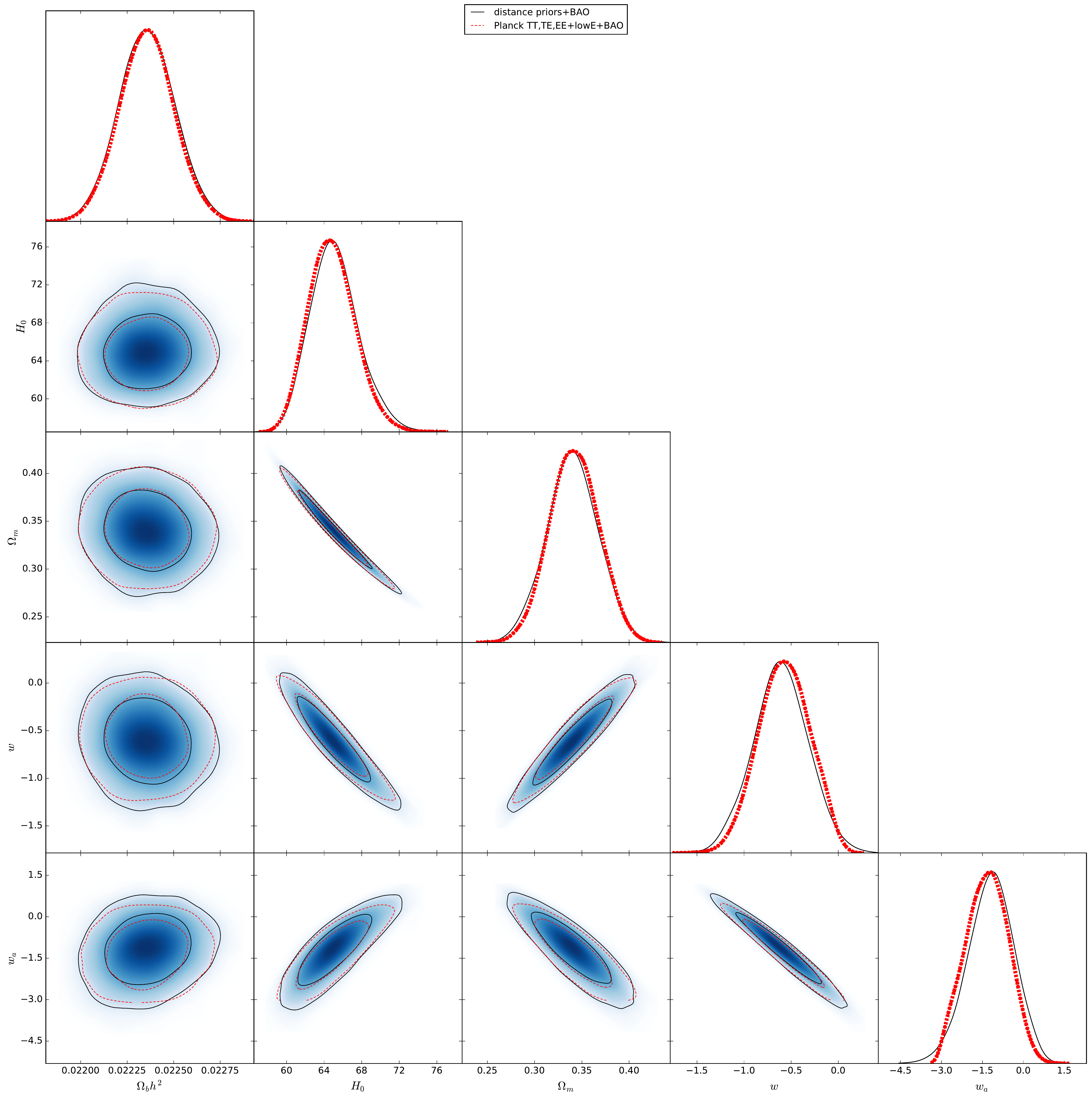}
\end{center}
\caption{Constraints on parameters in the CPL model. The $Planck$ 2018 chains gives the red dashed contours and the black solid ones are our results from distance priors for the base $\Lambda$CDM model.}
\label{fig:cpl}
\end{figure}

To constrain the equation of state of DE better, we combine the distance priors for the base $\Lambda$CDM model with the low redshift Baryon Acoustic Oscillation (BAO) measurements. We use the 6dFGS \cite{Beutler:2011hx}, SDSS-MGS \cite{Ross:2014qpa} and the final DR12 anisotropic BAO data \cite{Alam:2016hwk} at $z=0.106, 0.15, 0.38, 0.51, 0.61$.
Constraints on the parameters set $\{\Omega_m, H_0, \Omega_b h^2, w\}$ in the $w$CDM model and $\{\Omega_m, H_0, \Omega_b h^2, w,w_a\}$ in the CPL model are shown in Fig.~\ref{fig:w} and Fig.~\ref{fig:cpl} respectively. We can see that our results from distance priors for the base $\Lambda$CDM model and BAO measurements are consistent with those from $Planck$ 2018.

\section{Summary}
\label{sum}

In this work, we update the distance priors from the final release of the Planck Collaboration in the base $\Lambda$CDM model, the $w$CDM model, the $\Lambda$CDM$+\Omega_k$ model and the  $\Lambda$CDM$+A_\textrm{L}$ model. We give their mean values and the correlation matrices. Our new constraints on the distance priors are about $40\%$ tighter than those from $Planck$ 2015 TT$+$lowP \cite{Ade:2015rim}. Compared to our previous work \cite{Huang:2015vpa} based on the $Planck$ data release in 2015, our new results are slightly improved and $R$ in the base $\Lambda$CDM model gives the best improvement about $8\%$. 

We also check our results in the base $\Lambda$CDM model with the distance priors and constrain the related parameters in the $w$CDM model and the CPL model combining the low redshift BAO measurements. 
In all of these three DE models, we obtain quite similar constraints compared to $Planck$ 2018 release. 
It indicates that the distance priors derived from the base $\Lambda$CDM model can be used to replace the global fitting of full data released by $Planck$ in 2018 for the other DE models.

\vspace{5mm}
\noindent {\bf Acknowledgments}

We acknowledge the use of HPC Cluster of ITP-CAS. This work is supported by grants from NSFC (grant NO. 11335012, 11575271, 11690021, 11747601), Top-Notch Young Talents Program of China, and partly supported by the Strategic Priority Research Program of CAS and Key Research Program of Frontier Sciences of CAS.

\newpage

\begin{appendix}
\section{Note on using distance priors in the CosmoMC package}

In this appendix, we give a note about how to use the  distance priors in the CosmoMC package. Taking Fig.\ref{fig:lcdm} for example, the modified files are listed bellow.

1. Import the results about the distance priors in Table.\ref{distance priors}.

(i). Add the path of the distance priors parameter file named distance.ini in the CosmoMC input file test.ini.

\verb|~/cosmomc/test.ini|:
\begin{lstlisting}
...
DEFAULT(batch2/distance.ini)
...
\end{lstlisting}

(ii). Create the distance priors parameter file distance.ini.. The mean values of distances priors and $\Omega_b h^2$ should be included. 

\verb|~/cosmomc/batch2/distance.ini|:
\begin{lstlisting}
use_distance=T

redo_no_new_data =T
redo_add=T
redo_likeoffset = 0

prior_name = base_plikHM_TTTEEE_lowE
r=1.750235
la=301.4707
omegabh2=0.02235976

param[omegam]= 0.31 0 1 0.01 0.005
param[H0] = 68 20 100 0.1 0.1
param[omegak] = 0
param[mnu] = 0.06
param[w] = -1 
param[wa] = 0 
param[nnu] = 3.046
param[ombh2] = 0.0222 0.005 0.1 0.0001 0.0002
num_massive_neutrinos=1
distance_invcov_file = %DATASETDIR%Distance_invcov.txt
highL_theory_cl_template = %DATASETDIR%HighL_lensedCls.dat
\end{lstlisting}

(iii).  Create a file of the inverse matrix of the correlation matrix in Table.\ref{distance priors} and put it in the data folder of CosmoMC package. Actually, we use the inverse matrix before normalization here, instead of the inverse matrix of the nomalized one in Table.\ref{distance priors}.

\verb|~/cosmomc/data/Distance_invcov.txt|:
\begin{lstlisting}
#PLA 2018 base_plikHM_TTTEEE_lowE
#    R                l_a              a%omegabh2       
     94392.3971   -1360.4913   1664517.2916   
    -1360.4913   161.4349   3671.6180   
     1664517.2916   3671.6180   79719182.5162  
\end{lstlisting}

2. Set parameterization=background because we will run chains without the CMB calculation of the package and modified the background parameters as follows.

(i). \verb|~/cosmomc/source/driver.F90|:
\begin{lstlisting}
...
!call Setup%Config%SetTheoryParameterization(Ini, BaseParams%NameMapping,'theta')
call Setup%Config%SetTheoryParameterization(Ini, BaseParams%NameMapping,'background')
...
\end{lstlisting}

(ii). Specify the list of the background parameters.

\verb|~/cosmomc/paramnames/params_background.paramnames|:
\begin{lstlisting}
omegam        \Omega_m
H0            H_0
omegak        \Omega_K
mnu           \Sigma m_\nu		
w             w	
wa            w_a 
nnu           N_{eff}  
ombh2         \Omega_b h^2
\end{lstlisting}

(iii). Define the background parameters in the subroutine BK$\_$ParamArrayToTheoryParams in the same order of the file params$\_$background.paramnames. 

\verb|~/cosmomc/source/CosmologyParameterizations.f90|:
\begin{lstlisting}
subroutine BK_ParamArrayToTheoryParams(this, Params, CMB)
class(BackgroundParameterization) :: this
real(mcp) Params(:)
class(TTheoryParams), target :: CMB
real(mcp) omegam, h2
select type (CMB)
class is (CMBParams)
    omegam = Params(1)
    CMB%H0 = Params(2)
    CMB%omk = Params(3)
    CMB%omnuh2=Params(4)/neutrino_mass_fac*(standard_neutrino_neff/3)**0.75_mcp
    CMB%w =    Params(5)
    CMB%wa =    Params(6)
    CMB%nnu =    Params(7)
    CMB%h=CMB%H0/100
    h2 = CMB%h**2
    CMB%Yhe=0.24
    CMB%omnu = CMB%omnuh2/h2
    CMB%omegam =omegam
    CMB%omegamh2=CMB%omegam*h2
    CMB%ombh2=Params(8)
    CMB%omb=CMB%ombh2/h2
    CMB%omc= CMB%omegam - CMB%omnu - CMB%omb
    CMB%omch2 = CMB%omc*h2        
    CMB%zre=0
    CMB%tau=0
    CMB%omdmh2 = CMB%omch2+ CMB%omnuh2
    CMB%omdm = CMB%omdmh2/h2
    CMB%omv = 1- CMB%omk - CMB%omb - CMB%omdm
    CMB%nufrac=CMB%omnuh2/CMB%omdmh2
    CMB%reserved=0
    CMB%fdm=0
    CMB%iso_cdm_correlated=0
    CMB%Alens=1
end select
end subroutine BK_ParamArrayToTheoryParams
\end{lstlisting}

3. Add the likelihood of the distance priors and call it in the program.

(i). Create a new likelihood file named distance.f90 in the source folder. It is the main file including reading the mean values of distance priors and the inversed matrix mentioned before, as well as calculating  $\chi^2_{\rm distance\ priors}$.

\verb|~/cosmomc/source/distance.f90|:
\begin{lstlisting}
module distance
    use CosmologyTypes
    use MatrixUtils
    use Likelihood_Cosmology
    implicit none
    private
    
    type, extends(TCosmoCalcLikelihood) :: DistanceLikelihood
        real(mcp) :: R, la, omegabh2, ns
        real(mcp), allocatable, dimension(:,:) :: distance_invcov
    contains
    procedure :: LogLikeTheory => Distance_LnLike    
    end type DistanceLikelihood

    public DistanceLikelihood, DistanceLikelihood_Add
    contains

    subroutine DistanceLikelihood_Add(LikeList, Ini)
    class(TLikelihoodList) :: LikeList
    class(TSettingIni) :: ini
    Type(DistanceLikelihood), pointer :: this
    character(LEN=:), allocatable :: distance_invcov_file

    if (Ini%Read_Logical('use_distance',.false.)) then
        allocate(this)
        this%LikelihoodType = 'distance'
        this%name= Ini%Read_String('prior_name')
        this%R = Ini%Read_Double('r')
        this%la = Ini%Read_Double('la')
        this%omegabh2 = Ini%Read_Double('omegabh2')
        this%needs_background_functions = .true.
        call LikeList%Add(this)
        allocate(this%distance_invcov(3,3))
        this%distance_invcov=0
    end if 
    if (Ini%HasKey('distance_invcov_file')) then
        !write (*,*) 'start to read distance_invcov_file'
        distance_invcov_file  = Ini%ReadFileName('distance_invcov_file')
        call File%ReadTextMatrix(distance_invcov_file, this%distance_invcov)
        !write (*,*) this%distance_invcov
        !write (*,*) 'successly read distance_invcov_file'
    else
        write (*,*)'ERROR: distance_invcov_file'
    end if
    end subroutine DistanceLikelihood_Add

    real(mcp) function Distance_LnLike(this, CMB)
    use constants, only : c, const_pi
    Class(DistanceLikelihood) :: this
    Class(CMBParams) CMB
    real(mcp), dimension(3,3) :: invC
    real(mcp), dimension(3) :: x, d
    real(mcp) :: R, l_a, z_star, g_1, g_2

    g_1 = 0.0783D0*(CMB%ombh2)**(-0.238D0)/(1.0D0+39.5D0*(CMB%ombh2)**0.763D0)
    g_2 = 0.560D0/(1.0D0+21.1D0*(CMB%ombh2)**1.81D0)
    z_star = 1048.0D0*(1.0D0+0.00124D0*(CMB%ombh2)**(-0.738D0))*(1.0D0+g_1*(CMB%omegamh2)**g_2)
    R=CMB%H0*(CMB%omegam)**(0.5d0)*this%Calculator%AngularDiameterDistance(z_star)*(1.0D0+z_star)/c*1000.0D0
    l_a=const_pi/this%Calculator%CMBToTheta(CMB)    
    invC=this%distance_invcov
    d(1)=this%R
    d(2)=this%la
    d(3)=this%omegabh2
    !d(4)=this%ns
    x(1)=R
    x(2)=l_a
    x(3)=CMB%ombh2
    Distance_LnLike = DOT_PRODUCT((x-d),MATMUL(invC,(x-d)))/2.0d0
    end function  Distance_LnLike
    
    end module distance
\end{lstlisting}

(ii). Call the likelihood of distance priors in DataLikelihoods.f90.

\verb|~/cosmomc/source/DataLikelihoods.f90|:
\begin{lstlisting}
...
use distance
...
call DistanceLikelihood_Add(DataLikelihoods, Ini)
...
\end{lstlisting}

(iii). Modified the Makefile in the source folder and include the new lekelihood to make sure the program can be compiled successfully .

\verb|~/cosmomc/source/Makefile|:
\begin{lstlisting}
...
DATAMODULES = $(PLANCKLIKEFILES) $(OUTPUT_DIR)/mpk.o $(OUTPUT_DIR)/wigglez.o \
	$(OUTPUT_DIR)/bao.o $(SUPERNOVAE) $(SZ) $(OUTPUT_DIR)/supernovae.o $(OUTPUT_DIR)/HST.o  $(OUTPUT_DIR)/CMB.o $(OUTPUT_DIR)/CMBlikes.o $(OUTPUT_DIR)/ElementAbundances.o $(OUTPUT_DIR)/distance.o
...
$(OUTPUT_DIR)/distance.o: $(OUTPUT_DIR)/Likelihood_Cosmology.o
...
\end{lstlisting}

4. Sometimes people want or have to combine the distance priors with BAO measurements. In these cases, $r_s(z_d)$ is given by following code where the baryon drag epoch $z_d$ is given by Eisenstein \& Hu \cite{Eisenstein:1997ik}, instead of \verb|camb|. 

(i). \verb|~/cosmomc/source/bao.f90|:
\begin{lstlisting}
...
real(mcp) function get_rs_drag(this,CMB,Theory)
class(TBAOLikelihood) :: this
class(CMBParams) CMB
Class(TCosmoTheoryPredictions), target :: Theory
if (BAO_fixed_rs>0) then
    !this is just for use for e.g. BAO 'only' constraints
    get_rs_drag =  BAO_fixed_rs
else
    !get_rs_drag =  Theory%derived_parameters( derived_rdrag )
    get_rs_drag =  this%Calculator%distanceR(CMB)*148.92D0/153.017d0
end if
end function
...
\end{lstlisting}

(ii). \verb|~/cosmomc/source/Calculator_Cosmology.f90|:
\begin{lstlisting}
...
procedure :: ...
procedure :: distanceR
procedure :: ...
...
real(mcp) function distanceR(this, CMB)
class(TCosmologyCalculator) :: this
class(CMBParams) CMB
call this%ErrorNotImplemented('distanceR')
distanceR = 0
end function distanceR
...
\end{lstlisting}

(iii). \verb|~/cosmomc/source/Calculator_CAMB.f90|:
\begin{lstlisting}
...
procedure :: ...
procedure :: distanceR => CAMBCalc_distanceR
procedure :: ...
...
function CAMBCalc_distanceR(this, CMB) result(distanceR)
use ModelParams 
class(CAMB_Calculator) :: this
class(CMBParams) CMB
real(mcp) distanceR
distanceR = distanceOfR()
end function CAMBCalc_distanceR
...
\end{lstlisting}

(iv). \verb|~/cosmomc/camb/modules.f90|:
\begin{lstlisting}
module ModelParams
...
function distanceOfR()
real(dl) zdrag,adrag,atol
real(dl) distanceOfR
real(dl) obh2, omh2,b1,b2
real(dl) rombint
external rombint
obh2=CP%omegab*(CP%h0/100.0d0)**2
omh2=(CP%omegab+CP%omegac+CP%omegan)*(CP%h0/100.0d0)**2
b1     = 0.313D0*omh2**(-0.419D0)*(1.0D0+0.607D0*omh2**0.674D0)
b2     = 0.238D0*omh2**0.223D0
zdrag  = 1291.0D0*omh2**0.251D0*(1.0D0+b1*obh2**b2)/(1.0D0+0.659D0*omh2**0.828D0)
adrag = 1.0D0/(1.0D0+zdrag)
atol = 1e-6
distanceOfR=rombint(dsound_da,1d-8,adrag,atol)
end function distanceOfR

end module ModelParams
\end{lstlisting}

Now you can compile the program and run `./cosmomc test.ini' for test.

\end{appendix}




\begin{thebibliography}{99}
\frenchspacing

\bibitem{Riess:1998cb} 
  A.~G.~Riess {\it et al.} [Supernova Search Team],
  Astron.\ J.\  {\bf 116}, 1009 (1998)
  doi:10.1086/300499
  [astro-ph/9805201].
 

\bibitem{Perlmutter:1998np} 
  S.~Perlmutter {\it et al.} [Supernova Cosmology Project Collaboration],
  Astrophys.\ J.\  {\bf 517}, 565 (1999)
  doi:10.1086/307221
  [astro-ph/9812133].

\bibitem{Peebles:2002gy} 
  P.~J.~E.~Peebles and B.~Ratra,
  Rev.\ Mod.\ Phys.\  {\bf 75}, 559 (2003)
  doi:10.1103/RevModPhys.75.559
  [astro-ph/0207347].
  
\bibitem{Riess:2016jrr} 
  A.~G.~Riess {\it et al.},
  Astrophys.\ J.\  {\bf 826}, no. 1, 56 (2016)
  doi:10.3847/0004-637X/826/1/56
  [arXiv:1604.01424 [astro-ph.CO]].
 
\bibitem{Conley:2011ku} 
  A.~Conley {\it et al.} [SNLS Collaboration],
  Astrophys.\ J.\ Suppl.\  {\bf 192}, 1 (2011)
  doi:10.1088/0067-0049/192/1/1
  [arXiv:1104.1443 [astro-ph.CO]].
  
\bibitem{Suzuki:2011hu} 
  N.~Suzuki {\it et al.},
  Astrophys.\ J.\  {\bf 746}, 85 (2012)
  doi:10.1088/0004-637X/746/1/85
  [arXiv:1105.3470 [astro-ph.CO]].

\bibitem{Cole:2005sx} 
  S.~Cole {\it et al.} [2dFGRS Collaboration],
  Mon.\ Not.\ Roy.\ Astron.\ Soc.\  {\bf 362}, 505 (2005)
  doi:10.1111/j.1365-2966.2005.09318.x
  [astro-ph/0501174].
    
\bibitem{Sanchez:2006nj} 
  E.~Sanchez [DES Collaboration],
  AIP Conf.\ Proc.\  {\bf 878}, 213 (2006).
  doi:10.1063/1.2409089
  
\bibitem{Neveu:2017jkg} 
  J.~Neveu [LSST Dark Energy Science Collaboration],
  PoS EPS {\bf -HEP2017}, 045 (2017).
  doi:10.22323/1.314.0045
  
\bibitem{Gil-Marin:2015sqa} 
  H.~Gil-Marín {\it et al.},
  Mon.\ Not.\ Roy.\ Astron.\ Soc.\  {\bf 460}, no. 4, 4188 (2016)
  doi:10.1093/mnras/stw1096
  [arXiv:1509.06386 [astro-ph.CO]].
  
\bibitem{Alcock:1979mp} 
  C.~Alcock and B.~Paczynski,
  Nature {\bf 281}, 358 (1979).
  doi:10.1038/281358a0
   
\bibitem{Komatsu:2008hk} 
  E.~Komatsu {\it et al.} [WMAP Collaboration],
  Astrophys.\ J.\ Suppl.\  {\bf 180}, 330 (2009)
  doi:10.1088/0067-0049/180/2/330
  [arXiv:0803.0547 [astro-ph]].
  
\bibitem{Aghanim:2018eyx} 
  N.~Aghanim {\it et al.} [Planck Collaboration],
  arXiv:1807.06209 [astro-ph.CO].
 
\bibitem{Bean:2003fb} 
  R.~Bean and O.~Dore,
  Phys.\ Rev.\ D {\bf 69}, 083503 (2004)
  doi:10.1103/PhysRevD.69.083503
  [astro-ph/0307100].
  
\bibitem{Weller:2003hw} 
  J.~Weller and A.~M.~Lewis,
  Mon.\ Not.\ Roy.\ Astron.\ Soc.\  {\bf 346}, 987 (2003)
  doi:10.1111/j.1365-2966.2003.07144.x
  [astro-ph/0307104].
  
\bibitem{Li:2008cj} 
  H.~Li, J.~Q.~Xia, G.~B.~Zhao, Z.~H.~Fan and X.~Zhang,
  Astrophys.\ J.\  {\bf 683}, L1 (2008)
  doi:10.1086/591082
  [arXiv:0805.1118 [astro-ph]].
  
\bibitem{Dvali:2000xg} 
  G.~R.~Dvali and G.~Gabadadze,
  Phys.\ Rev.\ D {\bf 63}, 065007 (2001)
  doi:10.1103/PhysRevD.63.065007
  [hep-th/0008054].
  
\bibitem{Koyama:2006ef} 
  K.~Koyama,
  JCAP {\bf 0603}, 017 (2006)
  doi:10.1088/1475-7516/2006/03/017
  [astro-ph/0601220].
  
\bibitem{Bond:1997wr} 
  J.~R.~Bond, G.~Efstathiou and M.~Tegmark,
  Mon.\ Not.\ Roy.\ Astron.\ Soc.\  {\bf 291}, L33 (1997)
  doi:10.1093/mnras/291.1.L33
  [astro-ph/9702100].
  
\bibitem{Efstathiou:1998xx} 
  G.~Efstathiou and J.~R.~Bond,
  Mon.\ Not.\ Roy.\ Astron.\ Soc.\  {\bf 304}, 75 (1999)
  doi:10.1046/j.1365-8711.1999.02274.x
  [astro-ph/9807103].

\bibitem{Wang:2007mza} 
  Y.~Wang and P.~Mukherjee,
  Phys.\ Rev.\ D {\bf 76}, 103533 (2007)
  doi:10.1103/PhysRevD.76.103533
  [astro-ph/0703780].
  
\bibitem{Ade:2013zuv} 
  P.~A.~R.~Ade {\it et al.} [Planck Collaboration],
  Astron.\ Astrophys.\  {\bf 571}, A16 (2014)
  doi:10.1051/0004-6361/201321591
  [arXiv:1303.5076 [astro-ph.CO]].
  
\bibitem{Ade:2015xua} 
  P.~A.~R.~Ade {\it et al.} [Planck Collaboration],
  Astron.\ Astrophys.\  {\bf 594}, A13 (2016)
  doi:10.1051/0004-6361/201525830
  [arXiv:1502.01589 [astro-ph.CO]].

\bibitem{Planck:2006aa} 
  J.~Tauber {\it et al.} [Planck Collaboration],
  astro-ph/0604069.

\bibitem{Ade:2015rim} 
  P.~A.~R.~Ade {\it et al.} [Planck Collaboration],
  Astron.\ Astrophys.\  {\bf 594}, A14 (2016)
  doi:10.1051/0004-6361/201525814
  [arXiv:1502.01590 [astro-ph.CO]].


\bibitem{Huang:2015vpa} 
  Q.~G.~Huang, K.~Wang and S.~Wang,
  JCAP {\bf 1512}, no. 12, 022 (2015)
  doi:10.1088/1475-7516/2015/12/022
  [arXiv:1509.00969 [astro-ph.CO]].






  
  
  
  
  
  
  
  
  






\bibitem{Calabrese:2008rt} 
  E.~Calabrese, A.~Slosar, A.~Melchiorri, G.~F.~Smoot and O.~Zahn,
  Phys.\ Rev.\ D {\bf 77}, 123531 (2008)
  doi:10.1103/PhysRevD.77.123531
  [arXiv:0803.2309 [astro-ph]].
 
 
\bibitem{Lewis:2002ah} 
  A.~Lewis and S.~Bridle,
  Phys.\ Rev.\ D {\bf 66}, 103511 (2002)
  doi:10.1103/PhysRevD.66.103511
  [astro-ph/0205436].

   \bibitem{Hu:1995en}
  W.~Hu and N.~Sugiyama,
  Astrophys.\ J.\  {\bf 471}, 542 (1996)
  [astro-ph/9510117].
  
\bibitem{Beutler:2011hx} 
  F.~Beutler {\it et al.},
  Mon.\ Not.\ Roy.\ Astron.\ Soc.\  {\bf 416}, 3017 (2011)
  doi:10.1111/j.1365-2966.2011.19250.x
  [arXiv:1106.3366 [astro-ph.CO]].



\bibitem{Ross:2014qpa} 
  A.~J.~Ross, L.~Samushia, C.~Howlett, W.~J.~Percival, A.~Burden and M.~Manera,
  Mon.\ Not.\ Roy.\ Astron.\ Soc.\  {\bf 449}, no. 1, 835 (2015)
  doi:10.1093/mnras/stv154
  [arXiv:1409.3242 [astro-ph.CO]].

\bibitem{Alam:2016hwk} 
  S.~Alam {\it et al.} [BOSS Collaboration],
  Mon.\ Not.\ Roy.\ Astron.\ Soc.\  {\bf 470}, no. 3, 2617 (2017)
  doi:10.1093/mnras/stx721
  [arXiv:1607.03155 [astro-ph.CO]].

\bibitem{Eisenstein:1997ik} 
  D.~J.~Eisenstein and W.~Hu,
  Astrophys.\ J.\  {\bf 496}, 605 (1998)
  doi:10.1086/305424
  [astro-ph/9709112].


\end{thebibliography}
\end{document}